\documentclass[aps,prd,superscriptaddress,twocolumn,floatfix,longbibliography]{revtex4-2} 

\usepackage{graphicx,color}
\usepackage[colorlinks=true,linkcolor=blue,citecolor=blue,linkcolor=blue]{hyperref}%
\usepackage[normalem]{ulem}

\usepackage{amsmath,amssymb,amsfonts,bm}     
\usepackage[mathcal]{eucal}
\usepackage{dsfont}
\usepackage{array}
\usepackage{booktabs}

\allowdisplaybreaks[4]

\begin{document}


\title{Finite Temperature Casimir Effect of Scalar Field}


\author{Liang Chen} 
\email[Corresponding Email:]{slchern@ncepu.edu.cn}
\affiliation{School of Mathematics and Physics, North China Electric Power University, Beijing 102206, China}
\affiliation{Institute of Condensed Matter Physics, North China Electric Power University, Beijing 102206, China}
\affiliation{Hebei Key Laboratory of Physics and Energy Technology, North China Electric Power University, Baoding 071003, China}

\author{Yu-Jing Wang} 
\affiliation{School of Mathematics and Physics, North China Electric Power University, Beijing 102206, China}
\affiliation{Institute of Condensed Matter Physics, North China Electric Power University, Beijing 102206, China}

\author{Sheng-Yan Li} 
\affiliation{School of Mathematics and Physics, North China Electric Power University, Beijing 102206, China}
\affiliation{Institute of Condensed Matter Physics, North China Electric Power University, Beijing 102206, China}

\date{\today}

\begin{abstract}
We derive analytic expressions for the Helmholtz free energy, Casimir force, and Casimir entropy for both one-dimensional and three-dimensional scalar fields with Dirichlet boundary conditions at finite temperature. We investigate the negative Casimir entropy problem in these systems, as well as for a scalar field in the bulk of a three-dimensional sphere, and find that this issue arises under different regularization prescriptions with differing counterterms. We argue against introducing any counterterms for the thermal corrections to the Casimir effect and predict that the thermal-fluctuation-induced Casimir force becomes repulsive in the high-temperature regime --- for instance, when $aT/(\pi\hbar{v})>0.2419$ for three-dimensional scalar fields. 
\end{abstract}


\maketitle

\section{Introduction}

The uncertainty principle in quantum mechanics dictates that a harmonic oscillator in any quantum state cannot achieve absolute rest, possessing instead a nonzero ground state energy. This fundamental principle extends naturally to quantum field theory, where a free-space quantum field can be decomposed into a collection of harmonic oscillators characterized by distinct quantum numbers. Without regularization, the vacuum (ground state) energy of such a field exhibits divergence. Remarkably, modifying the field's boundary conditions can produce measurable physical effects, most notably the Casimir effect
\cite{Casimir1948PKNAW,Milonni1994QVacuumbook,Bordag2001PhysRep,Bordag2009Casimirbook,Klimchitskaya2009RMP}. While there is some controversy regarding the origin of the Casimir effect \cite{JaffeRL2005PRD}, we shall primarily consider it from the perspective of vacuum quantum fluctuations.
In this framework, fluctuations of the electromagnetic vacuum generate an attractive force between two parallel, neutral conducting plates, quantified at zero temperature by $-{\pi^2 c \hbar}/{(240a^4)}$, where $a$ represents the interplate separation, $\hbar$ and $c$ are the reduced Planck's constant and the speed of light in the vacuum, respectively. This interaction becomes dominant at micron-to-nanometer scales, profoundly impacting the design, functionality, and performance of micro- and nano-electromechanical devices. 
Recent decades have witnessed significant experimental progress across multiple frontiers of Casimir physics, including: repulsive Casimir forces
\cite{Munday2009MeasuredLR, Zhao2019StableCE, ZhangY2024NatPhys}, the critical Casimir effect \cite{Hertlein2008Nat, SoykaF2008PRL, Bonn2009PRL, Nguyen2013NatComm, Paladugu2016NatComm, Falko2022NatPhys}, the dynamical Casimir effect \cite{Wilson_2011, Lahteenmaki2013PNAS, Felicetti2014PRL, Vezzoli2018CommPhys}, the experimental constraints on nonstandard forces \cite{ChiaveriniJ2003PRL,MasudaM2009PRL,SushkovAO2011PRL,KlimchitskayaGL2012PRD,WangJ2016PRD,SedmikR2021Universe}, and the Casimir-Lifshitz torque \cite{Somers2018Nat}. Additionally, theoretical investigations have proliferated in various directions. The Casimir effect has been demonstrated to provide mechanisms for spontaneous compactification of extra spatial dimensions in theories beyond the Standard Model \cite{Mostepanenko1997Casimirbook}. Conversely, experimental measurement results can provide constraints on new theories \cite{PerivolaropoulosL2008PRD}. Repulsive Casimir interactions have been proposed in systems with special geometry \cite{LevinM2010PRL}, metamaterials \cite{ZhaoR2009PRL,SongG2017PRA}, topological matters \cite{Grushin2011PRL,Tse2012PRL,Rodriguez2014PRL,WilsonJ2015PRB}, or chiral media \cite{JiangQD2019PRB}. Furthermore, researchers have explored Casimir-like phenomena mediated by diverse quantum excitations beyond virtual photons, including gravitational fields \cite{Quach2015PRL}, massless fermions \cite{BellucciS2009PRD}, neutrinos \cite{Costantino2020JHEP}, phonons \cite{SchecterM2014PRL}, and magnons \cite{NakataK2023PRL}, significantly expanding the effect's conceptual and applicative horizons. 

Despite these significant advancements, a fundamental theoretical challenge persists in thermal Casimir physics --- recently termed the ``Casimir puzzle'' and ``Casimir conundrum'' by V. M. Mostepanenko \cite{Mostepanenko2021Universe} and extensively investigated over the past two decades \cite{Mostepanenko2006JPAMG,Mostepanenko2008JPA,Klimchitskaya2008JPAMT,Lamoreaux2012ARNPS,HartmannM2017PRL,LiuM2019PRB,Klimchitskaya2022IJMPA}. At its core lies an apparent incompatibility between the Drude model description of metals (including dc conductivity in dielectrics) and the predictions of Lifshitz theory. This discrepancy manifests most strikingly in two ways: (i) systematic deviations between Drude-model predictions and experimental measurements, and (ii) the model's violation of the Nernst heat theorem through its prediction of negative entropy \cite{BezerraVB2004PRA}. 
In this study, we revisit the longstanding problem of the finite-temperature Casimir effect for both three-dimensional (3D) and one-dimensional (1D) scalar fields  with Dirichlet boundary conditions. Our approach yields exact analytical expressions for the Helmholtz free energy, enabling rigorous evaluation of two commonly employed regularization schemes: (a) subtraction of free-space thermal fluctuations and (b) removal of divergent high-temperature terms. Our exact results reveal that both these two types of counterterms are inherently inconsistent, leading to the emergence of negative Casimir entropy in these regularization prescriptions.

We begin by reviewing the finite-temperature Casimir effect for electromagnetic (EM) fields confined between ideal metal plates. Crucially, the transverse electric (TE) and transverse magnetic (TM) modes obey distinct boundary conditions: while TE modes satisfy Dirichlet conditions, TM modes require special treatment of the $k_z=0$ component in the field quantization. This leads to the Helmholtz free energy per unit area,  
\begin{gather}
F_{\textrm{EM}}=-\frac{\pi^2\hbar{c}}{360a^3}-\sum_{l=1}^{\infty}\left[\frac{lT^2}{\hbar{c}a}\textrm{Li}_2\left(e^{-l\pi{\hbar}c/aT}\right)\right.\notag \\
\left.+\frac{T^3}{\pi\hbar^2c^2}\textrm{Li}_3\left(e^{-l\pi{\hbar}c/aT}\right)\right]-\frac{\zeta(3)T^3}{2\pi\hbar^2c^2}+\frac{\pi^2aT^4}{45\hbar^3c^3},  \label{eq_X1}
\end{gather}
where $T$ is the absolute temperature (with Boltzmann's constant $k_B=1$). Here $\textrm{Li}_n(z)$ and $\zeta(z)$ denote the polylogarithm and the Riemann zeta functions respectively. The terms represent: \\
\begin{itemize}
\item[1.] The first term: Zero-temperature quantum fluctuation contribution.
\item[2.] The summation over $l$: Contribution from thermal fluctuations of TE and TM modes, $l$ refers to the quantization of wavevector $k_z=l\pi/a$. 
\item[3.] The penultimate term: $k_z=0$ contribution of TM mode. 
\item[4.] The last term: the counterterm from free-space thermal radiation energy. 
\end{itemize}
Notably, the $k_z=0$ term exhibits unusual properties: it depends neither on plate separation $a$ nor contributes to the Casimir force, affecting only the entropy. 
Further investigations show that the Casimir entropy has the following asymptotic expressions, 
\begin{align}
S\left(\frac{aT}{\hbar{c}}\rightarrow0\right)&\approx\frac{3\zeta(3)}{2\pi\hbar^2{c^2}}T^2-\frac{4\pi^2}{45\hbar^3{c^3}}aT^3, \label{eq_X2} \\
S\left(\frac{aT}{\hbar{c}}\rightarrow\infty\right)&\approx\frac{\zeta(3)}{8\pi{a^2}}.  \label{eq_X3}
\end{align}
These results reveals two profound insights: (1) \emph{Entropy Saturation Puzzle}: Equation (\ref{eq_X3}) shows the Casimir entropy approaching a temperature-independent constant --- a thermodynamically unexpected behavior since entropy should grow indefinitely with accessible quantum states at fixed $a$; (2) \emph{Salvation Mechanism}: The $k_z=0$ TM mode term (first term in Eq. (\ref{eq_X2})) precisely cancels the negative entropy that would arise from the counterterm (second term in Eq. (\ref{eq_X2})). This raises our central question: What occurs when such a stabilizing term is absent, as in scalar fields with pure Dirichlet boundary conditions?

\section{Three-Dimensional Scalar Field}

\begin{figure}[tb]
	\centering
	\includegraphics[width=\linewidth]{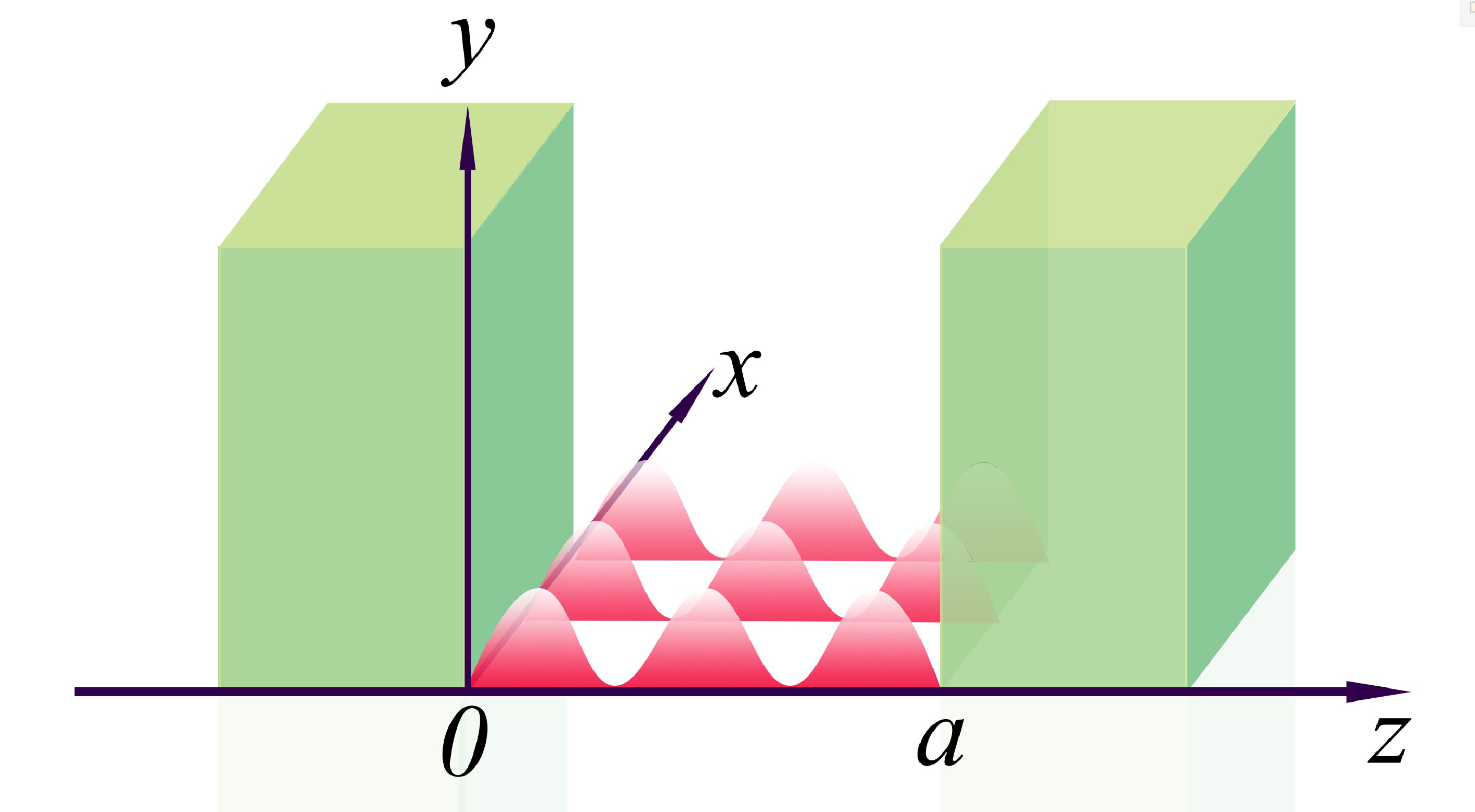}
	\caption{Illustration of the Casimir effect of 3D scalar field which has been constraint in the region $0<z<a$ with Dirichlet boundary condition. The waves in the region $0<z<a$ represent some of the scalar field configurations. 	
	\label{fig_X1}
	} 
\end{figure}

\begin{figure*}[tb]
	\centering
	\includegraphics[width=\linewidth]{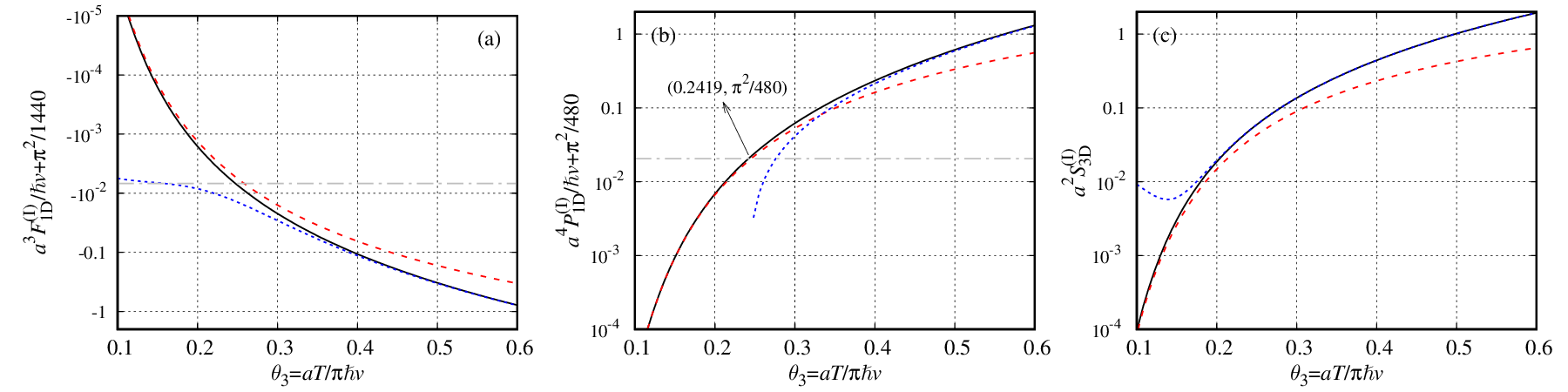}
	\caption{Helmholtz free energy (a), Casimir force (b), and Casimir entropy (c) as functions of dimensionless temperature $\theta_3=aT/\pi\hbar{v}$ for the 3D scalar field with Dirichlet boundary conditions. The solid black lines show the exact results. The dashed red lines show the low temperature approximation results. The dotted blue lines show the high-temperature approximations. In order to get clear presentation of the results, the Helmholtz free energy, Casimir force, and Casimir entropy are multiplied by $a^3/\hbar{v}$, $a^4/\hbar{v}$, and $a^2$, respectively. 
	\label{fig_X2}
	}
\end{figure*}

We first analyze the massless scalar field in three spatial dimensions, governed by the wave equation (see Fig. \ref{fig_X1} for illustration): 
\begin{equation}
	\left(\frac{\partial^2}{\partial{t^2}}-v^2\nabla^2\right)\phi(\bm{r},t)=0, \label{eq_X4}
\end{equation}
where $v$ represents the field propagation velocity. Through canonical quantization, we obtain the energy eigenvalues, 
\begin{equation}
	E_n(\bm{k}_{\parallel},l)=\left(n+\frac{1}{2}\right)\hbar{v}\sqrt{k_{\parallel}^2+\left(\frac{l\pi}{a}\right)^2}. \label{eq_X5}
\end{equation}
Here $\bm{k}_{\parallel}=(k_x,k_y)$ denotes the in-plane wavevector parallel to the boundaries, while the Dirichlet boundary conditions imposed at $z=0$ and $z=a$, 
\begin{equation}
	\phi(\bm{r}_{\parallel},0,t)=\phi(\bm{r}_{\parallel},a,t)=0,  \label{eq_X6}
\end{equation}
The integer $l=1$, $2,$, $...$ shown in Eq. (\ref{eq_X5}) indexes the discrete modes in the confined $z$-direction. The partition function of the system is given by, 
\begin{equation}
	Z_{3D}=\prod_{l,\bm{k}_{\parallel}}\sum_{n=0}^{\infty}e^{-{E_n(\bm{k}_{\parallel},l)/T}}, \label{eq_X7}
\end{equation}
From this, we derive the Helmholtz free energy per unit area through the standard relation $F=-T\log(Z)$,  
\begin{gather}
F_{3D}^{(\mathrm{I})}
=\sum_{l=1}^{\infty}\int_0^{\infty}\frac{k_{\parallel}dk_{\parallel}}{2\pi}\frac{1}{2}v\hbar\sqrt{k_{\parallel}^2+\left(\frac{l\pi}{a}\right)^2} \nonumber \\
+T\sum_{l=1}^{\infty}\int_0^{\infty}\frac{k_{\parallel}dk_{\parallel}}{2\pi}\log\left[1-e^{-\beta{v\hbar}\sqrt{k_{\parallel}^2+(l\pi/a)^2}}\right]. \label{eq_X8}
\end{gather}
The evaluation of this expression proceeds as follows: The first term is regularized using the Riemann zeta-function technique; For the second term, we perform the variable substitution, $u=\sqrt{k_{\parallel}^2+(l\pi/a)^2}$, yielding, 
\begin{equation}
	F_{3D}^{(\mathrm{I})}=-\frac{{\pi^2}v\hbar}{1440a^3}+\frac{T}{4\pi}\sum_{l=1}^{\infty}\int_{\left(\frac{l\pi}{a}\right)^2}^{\infty}\log\left(1-e^{-\frac{v\hbar}{T}\sqrt{u}}\right)du.  \label{eq_X9}
\end{equation}
The integration in the second term of Eq. (\ref{eq_X9}) can be expressed in terms of the polylogarithm function $\textrm{Li}_n(z)$, 
\begin{equation}
\frac{a^3}{v\hbar}F_{3D}^{(\mathrm{I})}=-\frac{{\pi^2}}{1440}
-\frac{\pi^2\theta^3}{2}\left(1-\frac{\partial}{\partial{s}}\right)\sum_{l=1}^{\infty}\mathrm{Li}_{3}\left(e^{-sl/\theta_3}\right)\Big|_{s=1}, \label{eq_X10}
\end{equation}
where we introduce the dimensionless temperature parameter $\theta_3=aT/{\pi}v\hbar$. The Casimir thermodynamic quantities like Casimir force ($P=-\partial{F}/\partial{a}$) and Casimir entropy ($S=-\partial{F}/\partial{T}$) per unit area are then obtained by direct differentiation,  
\begin{eqnarray}
\frac{a^4}{v\hbar}P_{3D}^{(\mathrm{I})}&=&-\frac{\pi^2}{480}-\frac{\pi^2\theta_3}{2}\sum_{l=1}^{\infty}l^2\log\left(1-e^{-l/\theta_3}\right), \label{eq_X11} \\
{a^2}S_{3D}^{(\mathrm{I})}&=&\frac{3\pi}{2}\sum_{l=1}^{\infty}\left[\theta_3^2\mathrm{Li}_3(e^{-l/\theta_3})+l\theta_3\mathrm{Li}_2(e^{-l/\theta_3}) \right.\nonumber \\
& & \left.
-\frac{l^2}{3}\log(1-e^{-l/\theta_3})\right]. \label{eq_X12}
\end{eqnarray}

Having derived the exact expressions, we now examine their asymptotic behavior in two characteristic limits. 
For the low-temperature regime where $\theta_3=aT/\pi\hbar{v}\ll1$ applies, the leading-order expansion of the polylogarithm function, 
\begin{equation}
	\mathrm{Li}_3(z)=\sum_{k=1}^{\infty}\frac{z^k}{k^n}, \label{eq_X13}
\end{equation}
yields simplified forms for the thermodynamic quantities. The Helmholtz free energy, Casimir force, and Casimir entropy reduces to, 
\begin{align}
	F_{3D}^{(\mathrm{I})}(aT\rightarrow0)&\approx-\frac{{\pi^2}\hbar{v}}{1440a^3}\left[1+720\left(\frac{aT}{\pi{\hbar}v}\right)^2e^{-\pi{\hbar{v}}/aT}\right], \label{eq_X14} \\
	P_{3D}^{(\mathrm{I})}(aT\rightarrow0)&\approx-\frac{{\pi^2}\hbar{v}}{480a^4}\left(1-240\frac{aT}{\pi\hbar{v}}e^{-\pi{\hbar{v}}/aT}\right), \label{eq_X15}
\\
	S_{3D}^{(\mathrm{I})}(aT\rightarrow0)&\approx\frac{\pi}{2a^2}\left(1+\frac{2aT}{\pi\hbar{v}}\right)e^{-\pi{\hbar}v/aT}. \label{eq_X16}
\end{align}
In the opposite high-temperature limit ($\theta_3=aT/\pi\hbar{v}\gg1$), the Casimir entropy asymptotically approaches, 
\begin{equation}
S_{3D}^{(\mathrm{I})}(aT\gg\pi\hbar{v})\approx\frac{\zeta(3)}{16\pi a^2}-\frac{3\zeta(3)}{4\pi\hbar^2v^2}T^2+\frac{2\pi^2}{45\hbar^3v^3}aT^3.  \label{eq_X17}
\end{equation}
Correspondingly, the Helmholtz free energy can be approximated by the following expression, 
\begin{gather}
F_{3D}^{(\mathrm{I})}(aT\gg\pi\hbar{v})\approx-\frac{\pi^2\hbar{v}}{1440a^3}-\frac{\zeta(3)}{16\pi a^2}T\nonumber \\
+\frac{\zeta(3)}{4\pi\hbar^2v^2}T^3-\frac{\pi^2a}{90\hbar^3v^3}T^4,  \label{eq_X18}
\end{gather}
and generates the temperature-dependent Casimir force given by, 
\begin{equation}
P_{3D}^{(\mathrm{I})}(aT\gg\pi\hbar{v})\approx-\frac{\pi^2\hbar{v}}{480a^4}-\frac{\zeta(3)}{8\pi a^3}+\frac{\pi^2}{90\hbar^3v^3}T^4.  \label{eq_X19}
\end{equation}

The high-temperature asymptotic expressions (Eqs. (\ref{eq_X17})-(\ref{eq_X19})) reveal a remarkable feature --- when the thermal energy dominates  ($aT\gg\pi{\hbar}v$), the Casimir force becomes repulsive and assumes a distance-independent value given by, 
\begin{equation}
	P_{3D}^{(\mathrm{I})}\approx\frac{\pi^2}{90\hbar^3v^3}T^4. \label{eq_X20}
\end{equation}
Figure \ref{fig_X2} displays the complete temperature dependence of (a) the Helmholtz free energy, (b) Casimir force, and (c) Casimir entropy, along with their low- and high-temperature approximations. The plots demonstrate excellent agreement between these approximations and the exact results in their respective temperature regimes. Notably, the transition between attractive and repulsive force regimes can be estimated analytically from the low-temperature approximation. This yields the critical ratio $aT/\pi\hbar{v}{\approx}W(240)=0.2454$, where $W$ denotes the Lambert W function. The analytical estimate shows close agreement with the numerically determined transition point at $0.2419$ (relative error $<1.5\%$).

We now examine how different regularization prescriptions affect the Casimir entropy. The $aT^4$ term in Eq. (\ref{eq_X18}) corresponds precisely to the thermal correction of the free-space Helmholtz energy density within the cavity $0<z<a$:  
\begin{equation}
	a\mathcal{F}_{\mathrm{free}}=a\int\frac{d\bm{k}}{(2\pi)^3}\left[\frac{1}{\beta}\log\left(1-e^{-\beta{v\hbar}k}\right)\right]=-\frac{\pi^2{aT^4}}{90v^3\hbar^3}. \label{eq_X21}
\end{equation}
A common regularization approach introduces a counterterm to subtract this free-space contribution, redefining the Helmholtz free energy as,
\begin{equation}
F_{3D}^{(\textrm{II})}=F_{3D}^{(\textrm{I})}-a\mathcal{F}_{\textrm{free}}.  \label{eq_X22}
\end{equation}
While this prescription eliminates the distance-independent repulsive force in the high-temperature limit ($aT/\pi\hbar{v}\gg1$), it creates an unphysical consequence --- the final term in Eq. (\ref{eq_X17}) also vanishes, leading to: 
\begin{equation}
S_{3D}^{(\mathrm{II})}(aT\gg\pi\hbar{v})\approx\frac{\zeta(3)}{16\pi a^2}-\frac{3\zeta(3)}{4\pi\hbar^2v^2}T^2<0,   \label{eq_X23}
\end{equation}
resulting in negative Casimir entropy at all temperatures, not just in the high-temperature regime.
Reference \cite{GeyerB2008EPJC} proposed a modified prescription that additionally subtracts $T^3$ and $T^2$ terms:  
\begin{equation}
F_{3D}^{(\textrm{III})}=F_{3D}^{(\textrm{I})}-a\mathcal{F}_{\textrm{free}}-\frac{\zeta(3)}{4\pi\hbar^2v^2}T^3.    \label{eq_X24}
\end{equation}
This alternative approach yields positive entropy that saturates at $\zeta(3)/16\pi a^2$ in the high-temperature limit, mimicking the EM field case. However, the critical question remains: Is this subtraction of $T^4$, $T^3$, and $T^2$ terms universally valid? To test this, we analyze the scalar field in a 3D spherical geometry.

\section{Scalar Field in three-dimensional sphere}

For a massless scalar field confined within a spherical cavity of radius $a$ under Dirichlet boundary conditions, the field quantization yields: 
\begin{equation}
\hat{\phi}(\bm{r},t)=\sum_{l,s}\left[\hat{a}_{l,s}e^{-i\omega_{l,s}{t}}\psi_{l,s}(\bm{r})+\hat{a}_{l,s}^{\dagger}e^{i\omega_{l,s}{t}}\psi^*_{l,s}(\bm{r})\right],   \label{eq_X25}
\end{equation}
where the eigenmodes $\psi_{l,s}(\bm{r})=j_l(ka)Y_{l,m}(\vartheta,\varphi)/\sqrt{N_{l,s}}$ satisfy, 
\begin{equation}
-\frac{v^2}{r^2}\left(\frac{\partial}{\partial{r}}r^2\frac{\partial}{\partial{r}}-\frac{L^2}{\hbar^2}\right)\psi_{l,s}(\bm{r})=v^2k_{l,s}^2\psi_{l,s}(\bm{r}).   \label{eq_X26}
\end{equation}
Here $Y_{l,m}$ are spherical harmonics with angular momentum quantum number $l$ ($L^2Y_{l,m}=l(l+1)Y_{l,m}$), while $s$ indexes the zeros of the spherical Bessel function, 
\begin{equation}
j_l(k_{l,s}a)=0.   \label{eq_X27}
\end{equation}
The corresponding eigenfrequencies are $\omega_{l,s}=vk_{l,s}$. 

The Hamiltonian diagonalizes to:  
\begin{equation}
\hat{H}=\sum_{l,s}\frac{\hbar{v}k_{l,s}}{2}\left(\hat{a}_{l,s}\hat{a}_{l,s}^{\dagger}+\hat{a}_{l,s}^{\dagger}\hat{a}_{l,s}\right),   \label{eq_X28}
\end{equation}
leading to the Helmholtz free energy decomposition:
\begin{equation}
F_{\textrm{sphere}}^{(\textrm{I})}=-T\log\left[\prod_{l,s}\sum_{n=0}^{\infty}e^{-\left(n+\frac{1}{2}\right)\hbar{v}k_{l,s}}\right],  \label{eq_X29}
\end{equation}
which naturally separates into zero-temperature quantum fluctuations and thermal contributions: 
$$F_{\textrm{sphere}}^{(\textrm{I})}=F_{\textrm{sphere}}^{(\textrm{I})}(T=0)+{\delta}F_{\textrm{sphere}}^{(\textrm{I})},$$ 
where the quantum fluctuation of the scalar field inside the sphere takes the standard form, 
\begin{equation}
F_{\textrm{sphere}}^{(\textrm{I})}(T=0)=\frac{\hbar{v}}{2}\sum_{l=0}^{\infty}\sum_{s=1}^{\infty}(2l+1)k_{l,s}.   \label{eq_X30}
\end{equation}
Different investigations show that \cite{NesterenkoVV1998PRD,BowersME1998PRD,CognolaG2001JPA,VALUYANMA2010IJMPA}, 
\begin{equation}
F_{\textrm{sphere}}^{(\textrm{I})}(T=0)=C\frac{\hbar{v}}{2a},   \label{eq_X31}
\end{equation}
where the dimensionless coefficient $C=0.00562$. 

Our primary interest lies in the thermal fluctuation component, 
\begin{equation}
{\delta}F_{\textrm{sphere}}^{(\textrm{I})}=T\sum_{l=0}^{\infty}\sum_{s=1}^{\infty}(2l+1)\log(1-e^{-\hbar{v}k_{l,s}/T}).    \label{eq_X32}
\end{equation}
Using the dimensionless temperature, $\theta_3=aT/\pi\hbar{v}$, the Casimir self-entropy $S=-\partial{F}/\partial{T}$ becomes, 
\begin{gather}
S_{\textrm{sphere}}^{(\textrm{I})}=-\sum_{l=0}^{\infty}\sum_{s=1}^{\infty}(2l+1)\log(1-e^{-\rho_{l,s}/\pi\theta_3})\notag\\
+\sum_{l=0}^{\infty}\sum_{s=1}^{\infty}(2l+1)\frac{\rho_{l,s}}{\pi\theta_3}\frac{1}{e^{\rho_{l,s}/\pi\theta_3}-1},   \label{eq_X33}
\end{gather}
where $\rho_{l,s}$ denote spherical Bessel zeros and the factor $2l+1$ accounts for magnetic quantum number degeneracy. 

\begin{figure}[tb]
	\centering
	\includegraphics[width=\linewidth]{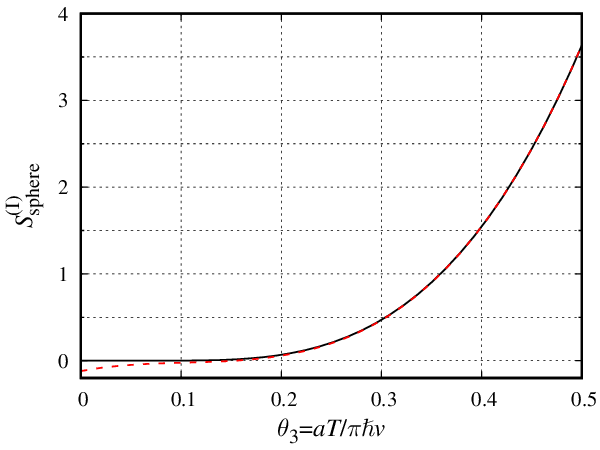}
	\caption{Solid line: Casimir self-entropy of scalar field in the 3D sphere as a function of dimensionless temperature multiplied by the radius. Dashed line: the high-temperature approximation given in Eq. (\ref{eq_X34}).
	\label{fig_X3}
	}
\end{figure}

Our numerical evaluation of the Casimir self-entropy via Eq. (\ref{eq_X33}) (see Supplementary Materials for computational details) considers all eigenmodes with angular and radial quantum numbers up to $l=s=1000$, encompassing $10^6$ distinct eigenenergies. In the high-temperature regime $\theta_3=aT/\pi\hbar{v}\gg1$, the entropy follows the polynomial expansion: 
\begin{equation}
S_{\textrm{sphere}}^{(\textrm{I})}=c_0+c_1\theta_3+c_2\theta_3^2+c_3\theta_3^3, \label{eq_X34}
\end{equation}
with coefficients $c_0=-0.1199$, $c_1=2.1689$, $c_2=-17.7931$, and $c_3=56.9711$. Notably, the cubic term precisely matches the blackbody radiation entropy for a free space volume $V=4\pi a^3/3$: 
\begin{equation}
\frac{4\pi a^3}{3}\mathcal{S}_{\textrm{free}}=\frac{8\pi^6}{135}\theta_3^3=56.9712\theta_3^3,    \label{eq_X35}
\end{equation}
where the entropy $\mathcal{S}_{\mathrm{free}}$ derives from the free Helmholtz free energy, $\mathcal{F}_{\mathrm{free}}$, i.e., $\mathcal{S}_{\mathrm{free}}=-\partial\mathcal{F}_{\mathrm{free}}/\partial{T}$. This high-precision agreement validates our numerical approach. 
Figure \ref{fig_X3} shows the high-temperature fitting of Casimir self-entropy using the polynomial expansion, Eq. (\ref{eq_X34}), revealing a fundamental dilemma:
\begin{itemize}
\item[1.] Subtracting only the blackbody term $c_3\theta_3^3$ yields negative entropy due to $c_2<0$. 
\item[2.] Removing all divergent terms ($c_1\theta_3$ through $c_3\theta_3^3$) again produces negative entropy from $c_0<0$. 
\item[3.] An ad hoc prescription keeping only $c_1\theta_3$ maintains positivity but lacks physical justification. 
\end{itemize}
This ambiguity underscores the need for universal regularization standards, motivating our subsequent analysis of the 1D scalar field case.

\section{One-Dimensional Scalar Field}

\begin{figure}[tb]
	\centering
	\includegraphics[width=\linewidth]{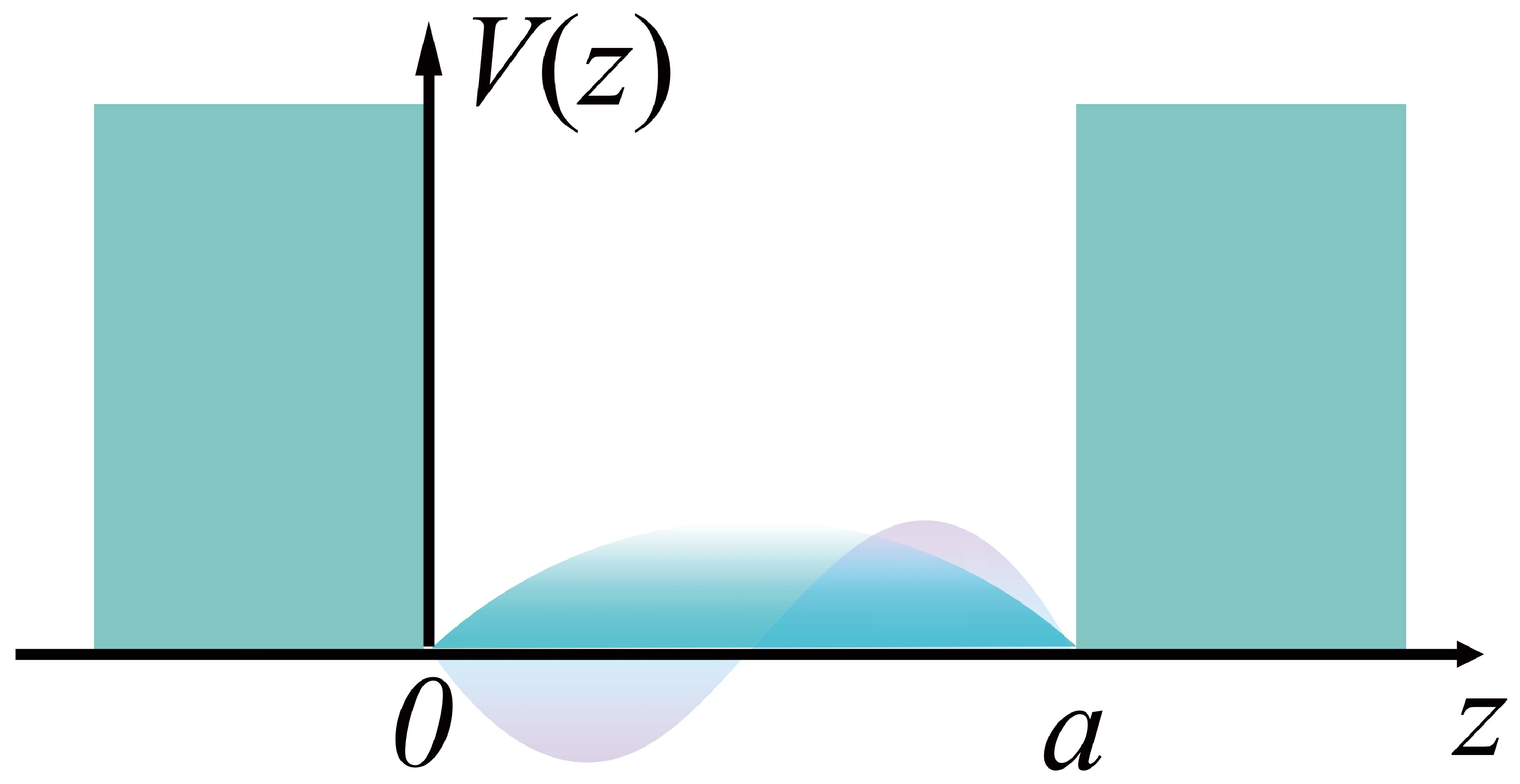}
	\caption{Illustration of the one-dimensional scalar field which has been constraint in the interval $[0,a]$ with Dirichlet boundary condition. The quantum fluctuation of this field can induce a Casimir interaction between the two boundaries. The waves in between the two boundaries show two possible modes of scalar field with eigenenergies $E_n(1)$ and $E_n(2)$, respectively. 
	\label{fig_X4}
	}
\end{figure}

\begin{figure*}[tb]
	\centering
	\includegraphics[width=\linewidth]{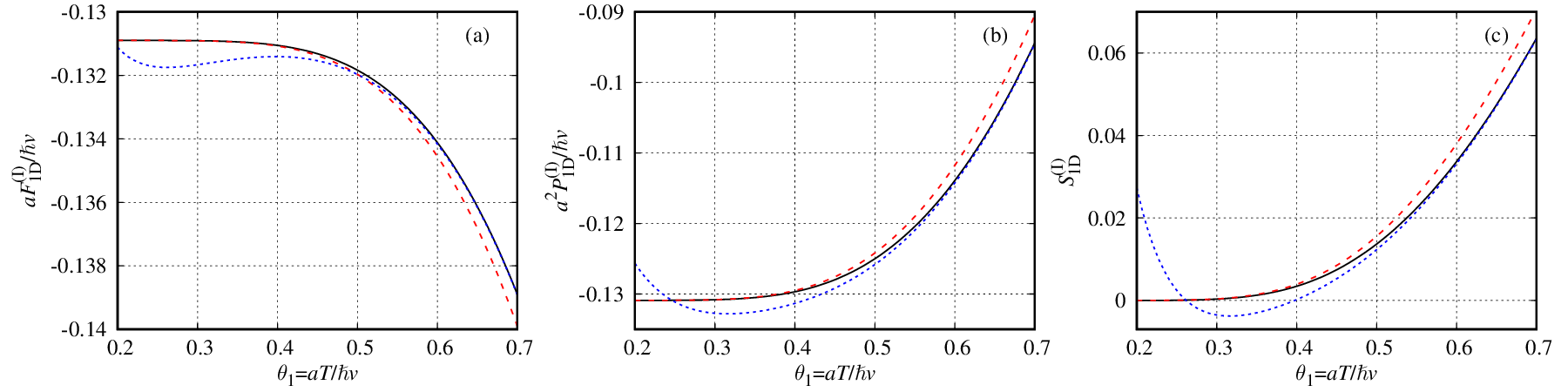}
	\caption{Helmholtz free energy (a), Casimir force (b), and Casimir entropy (c) as functions of dimensionless temperature $\theta_1=aT/\hbar{v}$ for the 1D scalar field with Dirichlet boundary conditions. The solid black lines show the exact results. The dashed red lines show the low temperature approximation results. The blue dotted lines show the high-temperature approximations. In order to get dimensionless results, the Helmholtz free energy and Casimir force are multiplied by $a/\hbar{v}$ and $a^2/\hbar{v}$, respectively. 
	\label{fig_X5}
	}
\end{figure*}

Consider a massless scalar field in one spatial dimension (Fig. \ref{fig_X4}) governed by the wave equation: 
\begin{equation}
	\left( \frac{\partial^2}{\partial{t}^2} - v^2 \frac{\partial^2}{\partial{x}^2} \right) \phi(x,t) = 0, \label{eq_X36}
\end{equation}
where $v$ denotes the field propagation velocity. Under Dirichlet boundary conditions, $\phi(0,t)=\phi(a,0)=0$, the field quantization yields discrete eigenenergies:
\begin{equation}
	E_n(l)=\left(n+\frac{1}{2}\right)\hbar{v}\frac{l\pi}{a},  \label{eq_X37}
\end{equation}
where $n=0, 1, 2, ...$ represents the particle number index, while $l=1, 2, 3, ...$ quantizes the wavevector as $k=l\pi/a$. 
The finite-temperature partition function evaluates to: 
\begin{equation}
	Z=\prod_{l=1}^{\infty}\sum_{n=0}^{\infty}e^{-{E_n(l)/T}}=\prod_{l=1}^{\infty}\frac{\exp\left(-{l\pi{v}\hbar}/{2aT}\right)}{1-\exp\left(-{l\pi{v}\hbar}/{aT}\right)}, \label{eq_X38}
\end{equation} 
from which we derive the Helmholtz free energy $F=-T\log(Z)$: 
\begin{equation}
	F_{1D}^{(\mathrm{I})}=\sum_{l=1}^{\infty}\left\{\frac{l\pi{v}\hbar}{2a}+T\log\left[1-\exp\left(-\frac{l\pi{v}\hbar}{aT}\right)\right]\right\}. \label{eq_X39}
\end{equation}
Applying  Riemann $\zeta$-function regularization ($\sum_{l=1}^{\infty}l=\zeta(-1)=-1/12$) to the first term in Eq. (\ref{eq_X39}) gives, 
\begin{equation}  
	F_{1D}^{(\mathrm{I})}=-\frac{\pi{v}\hbar}{24a}+T\sum_{l=1}^{\infty}\log\left[1-\exp\left(-\frac{l\pi{v}\hbar}{aT}\right)\right], \label{eq_X40} 
\end{equation}
while the second term summation produces, 
\begin{equation} 
	\sum_{l=1}^{\infty}\log\left[1-\exp\left(-\frac{l\pi{v}\hbar}{aT}\right)\right]=\frac{\pi{v}\hbar}{24aT}+\log\left[\eta\left(\frac{iv\hbar}{2aT}\right)\right],   \label{eq_X41} 
\end{equation}
in terms of the Dedekind $\eta$-function defined on the upper half complex plane ($\mathrm{Im}[z]>0$): 
\begin{equation}
	\eta(z)=q^{1/24}\prod_{n=1}^{\infty}(1-q^n), ~~~ (q=e^{2i\pi{z}}). \label{eq_X42}
\end{equation}
This function possesses the modular transformation property:
\begin{equation} 
	\eta(z)=\sqrt{\frac{i}{z}}\eta\left(-\frac{1}{z}\right).  \label{eq_X43}
\end{equation}
Remarkably, substituting Eq. (\ref{eq_X41}) into Eq. (\ref{eq_X40}) leads to exact cancellation of the $a^{-1}$ terms, yielding the simplified Helmholtz free energy: 
\begin{equation}
	F_{1D}^{(\mathrm{I})}=T\log\left[\eta\left(\frac{iv\hbar}{2aT}\right)\right].   \label{eq_X44}
\end{equation}
The Casimir force $P=-\partial{F}/\partial{a}$ and the Casimir entropy $S=-\partial{F}/\partial{T}$ can be calculated directly through differentiation, 
\begin{gather}
P_{1D}^{(\mathrm{I})}=\frac{iv\hbar}{2a^2}\frac{\eta^{\prime}\left({iv\hbar}/{2aT}\right)}{\eta\left({iv\hbar}/{2aT}\right)},
 \label{eq_X45} \\
S_{1D}^{(\mathrm{I})}=-\log\left[\eta\left(\frac{iv\hbar}{2aT}\right)\right]+\frac{iv\hbar}{2aT}\frac{\eta^{\prime}\left({iv\hbar}/{2aT}\right)}{\eta\left({iv\hbar}/{2aT}\right)}.   \label{eq_X46} 
\end{gather}
Introducing the dimensionless temperature $\theta_1={aT}/{v\hbar}$, we obtain the compact forms: 
\begin{gather}
\frac{a}{v\hbar}F_{1D}^{(\mathrm{I})}=\theta_1\log\left[\eta\left(\frac{i}{2\theta_1}\right)\right],   \label{eq_X47} \\
\frac{a^2}{v\hbar}P_{1D}^{(\mathrm{I})}=\frac{i}{2}\frac{\eta^{\prime}(i/2\theta_1)}{\eta(i/2\theta_1)},   \label{eq_X48} \\
S_{1D}^{(\mathrm{I})}=-\log\left[\eta\left(\frac{i}{2\theta_1}\right)\right]+\frac{i}{2\theta_1}\frac{\eta^{\prime}(i/2\theta_1)}{\eta(i/2\theta_1)}.   \label{eq_X49}
\end{gather}

Having established the exact solutions for the 1D scalar field, Eqs. (\ref{eq_X44})-(\ref{eq_X46}), we now examine their asymptotic limits to reveal the underlying thermodynamic behavior. In the low temperature region ($\theta_1=aT/{v}\hbar\ll1$), the Dedekind $\eta$ function can be approximated as \cite{Polchinski1998string,Green1987superstring}, 
\begin{equation}
\eta(z)=A\frac{q^{1/24}}{\sqrt{1-q}}\exp\left[-\frac{\pi^2}{6(1-q)}\right], (q=e^{2i\pi{z}}), \label{eq_X50} 
\end{equation}
where $A$ is a coefficient need to be determined. We choose $A=e^{\pi^2/6}$ by considering that $S(T\rightarrow0)=0$. Using this approximation, we can get the Helmholtz free energy, Casimir force, and Casimir entropy in the low temperature limit, 
\begin{gather}
\frac{a}{v\hbar}F_{1D}^{(\mathrm{I})}=-\frac{\pi}{24}-\frac{\theta_1}{2}\left(\frac{\pi^2}{3}-1\right)e^{-\pi/\theta_1} , \label{eq_X51} \\
\frac{a^2}{v\hbar}P_{1D}^{(\mathrm{I})}=-\frac{\pi}{24}+\frac{\pi}{2}\left(\frac{\pi^2}{3}-1\right)e^{-\pi/\theta_1}, \label{eq_X52} \\
S_{1D}^{(\mathrm{I})}=\frac{\pi}{2}\left(\frac{\pi^2}{3}-1\right)\left(\frac{\pi}{\theta_1}+1\right)e^{-\pi/\theta_1}. \label{eq_X53} 
\end{gather}
In the high temperature region ($\theta_1=aT/v\hbar\gg1$), applying the modular transformation property of the Dedekind $\eta$-function, Eq. (\ref{eq_X43}),  to the Helmholtz free energy,  Eq. (\ref{eq_X44}), yields the equivalent expression: 
\begin{equation}
	F_{1D}^{(\mathrm{I})}=T\log\left[\sqrt{2\theta_1}\eta\left({2i\theta_1}\right)\right]. \label{eq_X54}
\end{equation}
Substituting the asymptotic form of the $\eta$-function, Eq. (\ref{eq_X50}), into this transformed expression, we obtain the high-temperature approximations for the thermodynamic quantities: 
\begin{gather}
\frac{a}{v\hbar}F_{1D}^{(\mathrm{I})}=
\frac{\log(2\theta_1)}{2}\theta_1-\frac{\pi}{6}\theta^2-\frac{1}{2}\left(\frac{\pi^2}{3}-1\right)e^{-4\pi\theta_1}, \label{eq_X55} \\
\frac{a^2}{v\hbar}P_{1D}^{(\mathrm{I})}=-\frac{\theta_1}{2}+\frac{\pi}{6}\theta^2-2\pi\left(\frac{\pi^2}{3}-1\right)\theta_1^2e^{-4\pi\theta_1}, \label{eq_X56} \\
S_{1D}^{(\mathrm{I})}=-\frac{1+\log(2\theta_1)}{2}+\frac{\pi}{3}\theta_1\nonumber\\
-\frac{1}{2}\left(\frac{\pi^2}{3}-1\right)(4\pi\theta_1-1)e^{-4\pi\theta_1}. \label{eq_X57}
\end{gather}
Numerical evaluation confirms excellent agreement between these asymptotic approximations and the exact solutions (Fig. \ref{fig_X5}). Most notably, the Casimir entropy maintains positive values at all temperatures when no counterterms are introduced in this system.

\begin{figure}[tb]
	\centering
	\includegraphics[width=\linewidth]{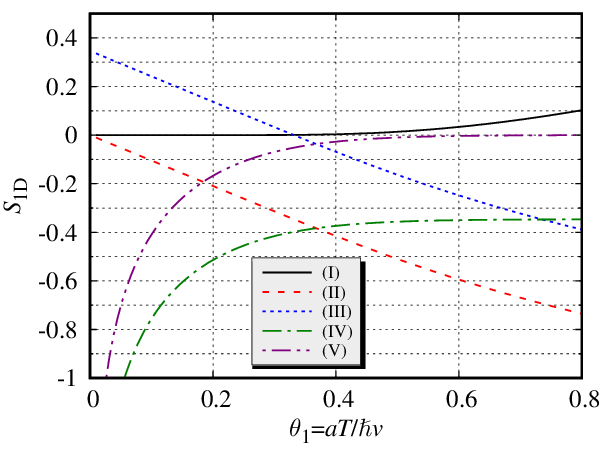}
	\caption{Casimir entropy with different counterterms for the 1D scalar field with Dirichlet boundary condition. Details of the counterterms are given in Table \ref{tab_X1}. The solid black line shows our proposal without counterterms for the thermal fluctuation of the scalar field. 
	\label{fig_X6}
	}
\end{figure}

\begin{table}[tb] 
\renewcommand\arraystretch{1.5}
\caption{Summarize of different regularization prescriptions. \label{tab_X1} }
\begin{tabular}{p{0.22\columnwidth}<{\centering}|p{0.34\columnwidth}<{\centering}|p{0.38\columnwidth}<{\centering}}
    \hline\hline
    {Prescription (ind.)} & Counterterm in Helmholtz Free Energy  $\frac{a}{\hbar{v}}\left(F_{1D}^{(\textrm{ind.})}-F_{1D}^{(\textrm{I})}\right)$ & Modification of Casimir Entropy $\left(S_{1D}^{(\textrm{ind.})}-S_{1D}^{(\textrm{I})}\right)$ \\
    \hline
    (\textrm{I})  & - & -  \\
    (\textrm{II}) & $\frac{\pi}{6}\theta_1^2$ & $-\frac{\pi}{3}\theta_1$  \\
    (\textrm{III}) & $\frac{\pi}{6}\theta_1^2-\frac{1}{2}\log(2)\theta_1$ & $-\frac{\pi}{3}\theta_1+\frac{1}{2}\log(2)$  \\
    (\textrm{IV}) & $\frac{\pi}{6}\theta_1^2-\frac{1}{2}\theta_1\log(\theta_1)$ &  $-\frac{\pi}{3}\theta_1+\frac{1}{2}[\log(\theta_1)+1]$ \\
    (\textrm{V}) & $\frac{\pi}{6}\theta_1^2-\frac{1}{2}\theta_1\log(2\theta_1)$ & $-\frac{\pi}{3}\theta_1+\frac{1}{2}[\log(2\theta_1)+1]$  \\
    \hline\hline
\end{tabular}
\end{table}

We now examine the free-space energy density. The second term in Eq. (\ref{eq_X40}), gives the thermal fluctuation contribution to the free energy density in unbounded space, 
\begin{equation}
	a\mathcal{F}_{\mathrm{free}}=aT\int_{-\infty}^{\infty}\frac{dk}{2\pi}\log\left(1-e^{-{v\hbar}k/T}\right)=-\frac{\pi}{6}\frac{aT^2}{v\hbar}, \label{eq_X58}
\end{equation}
Adopting this term as a counterterm removes the leading divergence in Eq. (\ref{eq_X55}) at high temperatures: 
\begin{equation}
	F_{1D}^{(\mathrm{II})}=F_{1D}^{(\mathrm{I})}-a\mathcal{F}_{\mathrm{free}}=T\log\left[\eta\left(\frac{iv\hbar}{2aT}\right)\right]+\frac{\pi}{6}\frac{aT^2}{v\hbar}. \label{eq_X59}
\end{equation}
This modification eliminates the dominant positive Casimir entropy contribution, yielding:
\begin{gather}
S_{1D}^{(\mathrm{II})}(\theta_1\gg1)=-\frac{\partial{F_{1D}^{(\mathrm{II})}}}{\partial{T}} \nonumber\\
\approx-\frac{1+\log(2\theta_1)}{2}-\frac{1}{2}\left(\frac{\pi^2}{3}-1\right)(4\pi\theta_1-1)e^{-4\pi\theta_1}, \label{eq_X60}
\end{gather}
The Casimir entropy becomes negative to leading order in $e^{-4\pi\theta_1}$ and remains negative at all temperatures when computed via Eq. (\ref{eq_X60}). We have tested alternative regularization prescriptions (see Tab. \ref{tab_X1} and Fig. \ref{fig_X6} for details). All prescriptions produce negative Casimir entropy in some or all temperature ranges. This highlights a persistent challenge: using the free-space thermal fluctuation as a counterterm necessarily introduces negative Casimir entropy. For 1D scalar field with Dirichlet boundary conditions, no physically justified counterterm prescription avoids this outcome.

\section{Summary}

We have systematically investigated the finite-temperature Casimir effect for scalar fields in one and three spatial dimensions. For the 3D scalar field confined between parallel plates ($0<z<a$) with Dirichlet boundary conditions, our analysis shows that:
\begin{itemize}
\item[1.] Using only the free-space thermal radiation energy as a counterterm leads to negative Casimir entropy. 
\item[2.] Including additional $T^3$ and $T^2$ terms in the counterterm prescription restores positive entropy, with the high-temperature limit approaching $\zeta(3)/(16\pi{a^2})$. 
\item[3.] For the spherical geometry, high-precision numerical calculations reveal that even this extended prescription fails, yielding a negative high-temperature entropy limit ($\sim-0.12$).
\end{itemize}

In the 1D case with exact solutions, all four tested counterterm schemes (Table \ref{tab_X1}) produced negative entropy in some temperature regimes. These findings extend previous studies of negative entropy \cite{Milton2017PRD,Bordag2018PRD,Milton2019PRD,LiY2021Entropy}, demonstrating that even for simple Dirichlet boundary conditions with exact solutions, conventional counterterm approaches face fundamental challenges. 

Our results raise important questions about standard regularization prescriptions:
\begin{itemize}
\item[1.] Counterterms, while mathematically useful for removing divergences, may lack physical justification when finite results exist without them.
\item[2.] The persistent negative entropy problem suggests deeper issues in thermal Casimir physics that warrant re-examination of standard approaches like the Lifshitz formula and quantum field theory approach.
\end{itemize}

Notably, when avoiding problematic counterterms (prescriptions used in this work), our theory predicts a sign reversal of the Casimir force at $\theta=aT/\pi{c}{\hbar}{\gtrsim}0.2419$. For $T=300  \mathrm{~K}$ and $v=c$, this corresponds to $a{\gtrsim}5.8\mathrm{~{\mu}m}$ --- within reach of current state-of-art experimental capabilities \cite{SushkovAO2011NatPhys}. This repulsive regime, where thermal radiation pressure dominates quantum fluctuations, provides a testable signature of our approach.

\section{Acknowledgments}		
The authors are grateful for financial support from the National Natural Science Foundation of China (Grant No. 12174101) and the Fundamental Research Funds for the Central Universities (Grant No. 2022MS051).

\bibliography{references.bib}

\end{document}